\documentclass[11pt]{article}
\usepackage{epsfig,enumerate,verbatim}
\usepackage{amsmath}
\usepackage{amsfonts}
\usepackage{amssymb}
\usepackage{amstext}
\usepackage{amsthm}
\usepackage{undertilde} % command is \utilde
\usepackage{dsfont}
\usepackage{graphicx}
\usepackage{mathrsfs}
\begin{document}        
\title{\bf{Scattering and Bound States of a Deformed Quantum Mechanics}}

\author{Chee-Leong Ching\footnote{Email: phyccl@nus.edu.sg} and Rajesh R. Parwani\footnote{Email: parwani@nus.edu.sg}}

\date{6 July 2012}

\maketitle

\begin{center}
{Department of Physics,\\}
{National University of Singapore,\\}
{Kent Ridge,\\}
{Singapore.}
\end{center}

\vspace{1.5cm}

\begin{abstract}
\noindent We construct the exact position representation of a deformed quantum mechanics which exhibits an intrinsic maximum momentum and use it to study problems such as a particle in a box and scattering from a step potential, among others. In particular, we show that unlike usual quantum mechanics, the present deformed case delays the formation of bound states in a finite potential well. In the process we also highlight some limitations and pit-falls of low-momentum or perturbative treatments and thus resolve two puzzles occurring in the literature.

\end{abstract}

\section{Introduction}

Modified quantum mechanical commutation relations have been extensively studied as an effective means of encoding potential gravitational or other effects; see Refs.\cite{sabine,optics} for a complete list of references. 

In our previous paper \cite{spectra} we investigated in detail a deformed quantum mechanics described by a modified commutation relation (MCR) exhibiting maximum momentum. In one dimension, the relevant modified Heisenberg algebra\footnote{We choose units where $\hbar$=$2M$=1, $M$ the particle mass.} is

\begin{equation}
[X,P] = i f(P)  \, , \label{mod}
\end{equation}
with $f(p) =1$ in usual quantum mechnics. Intrinsic maximum momentum arises when $f(P)$ has a singularity \cite{PedramMax, spectra} or a zero \cite{dsr,das,mig,jiz,cutoff,spectra} at some $P=P_0$. One effect of maximum momentum is that the spectrum of bound states terminates at finite energy even for potentials such as the harmonic oscillator \cite{spectra}; this is in contrast to MCR's which exhibit instead a minimum position uncertainty \cite{kempf,snyder}. 

In \cite{spectra} we focussed on the class of MCR's defined by 
\begin{equation}
f(P) = 1-2\alpha P + q\alpha^2 P^2 \label{class}
\end{equation} 
where $\alpha >0$ and $q$ are real parameters\footnote{Although in \cite{spectra} we also studied higher space dimensions, here we restrict ourselves to the one-dimensional subspace}. For $q \le 1$ Eq.(\ref{class}) has a zero and hence (in momentum space) Eq.(\ref{mod}) implies an intrinsic maximum momentum. For $q>1$ there is no intrinsic maximum momentum but rather a minimum position uncertainty. 

Note that in the limits $\alpha \to 0$ and $q\alpha^2 \to \pm \beta^2$, Eq.(\ref{class}) describes the Snyder (or string-motivated) \cite{kempf,snyder} and Anti-Snyder \cite{mig} algebra's.  The first case, with $+\beta^2$, allows for minimum position uncertainty while the $-\beta^2$ case exhibits maximum momentum. 

The study of Eq.(\ref{mod}) in Ref.\cite{spectra} used the momentum representation
\begin{eqnarray}
P&=&p \, , \label{momrep1}\\
X&=& i f(p) \frac{\partial}{\partial p}  \,  \label{momrep} \, , \\
\left[x,p\right] &= & i  \, , \label{canon}
\end{eqnarray}
with $ x= i \frac{\partial}{\partial p}$. In this paper we will instead adopt the position representation
\begin{eqnarray}
P&=& P(p) \, , \label{postrep1}\\
X&=& x  \, , \label{postrep}  \\
\left[x,p\right] &= & i  \, , \label{canon} \\
p &=& -i \frac{\partial}{\partial x} \, ,  \label{postrep2}
\end{eqnarray}
where the functional form of the operator $P(p)$ will be determined exactly in Sect.(2). 

The discussion of certain problems, such as scattering from a barrier or a particle in a square well, is much easier in the position representation. Except for the Snyder limit \cite{Pedramx}, previous studies \cite{das,brau,das-rel,ghosh, Pedram2} of the position representation for the class (\ref{class}) have used truncated  low-momentum expansions of the form
\begin{equation}
P = p ( 1- 2 \alpha p + O(p^2) \ ) \, . \label{lowmom}
\end{equation}
 While such an approach is adequate for many practical situations, it has a number of limitations and pit-falls.
 
 For example, as already discussed in \cite{spectra}, an expansion of the form (\ref{lowmom}) cannot uncover the precise effects of an intrinsic  maximum momentum which occur at $P \sim 1/\alpha$. Indeed, even if there is no maximum momentum, as when $q>1$, the reliability of an expansion such as (\ref{lowmom}) requires that successive terms do {\it not} exceed previous terms so that a truncation gives accurate approximations. So, comparing the first two terms in (\ref{lowmom}),  we need at least $|2\alpha p^2/p| < 1$, that is\footnote{Here and below to be interpreted as constraints on the eigen or mean values.} 
\begin{equation}
 |p| < 1/( 2 \alpha )  \, , \label{const}
\end{equation}
though, for better accuracy, it is usual to require the stronger condition whereby ``$<$" in the last inequality is replaced by ``$\ll$". 
 
 Errors may arise if one forgets to implement the constraint (\ref{const}) when using the truncation (\ref{lowmom}). As a toy illustration, consider first the algebraic equation
 \begin{equation}
 {1 \over 1-\epsilon \nu} = 1 \label{toy}
 \end{equation}
 where $\nu$ is an unknown and $\epsilon$ a small parameter. The unique exact solution is $\epsilon \nu =0$. However if the left-side of (\ref{toy}) is first expanded to quadratic order and truncated, the resulting equation $1+ \epsilon \nu + (\epsilon \nu)^2 =1$ has an extra solution $\epsilon \nu = -1$. The second solution is spurious and can be discarded as it does not satisfy the consistency condition (analogous to (\ref{const})  $|\epsilon \nu/1| <1$.  

Notice that knowledge of the exact solution was not required in order to eliminate the spurious solution: The simple consistency criterion was sufficient. Another toy model, closer to the concerns of this paper, is given by the non-local differential equation
\begin{equation}
 {1 \over 1-\epsilon \partial_u} v(u) = v(u) \label{toy2}
\end{equation}
whose unique exact solution is clearly $v(u)=$ constant . On the other hand expanding the left-side to quadratic order and truncating gives a second-order differential equation $v + \epsilon \partial_u v + (\epsilon \partial_u)^2 v = v$ which has the extra solution $v(u) \propto e^{-u/\epsilon}$. But the consistency condition $|\epsilon \partial_u v / v | <1$ shows that the extra solution should be discarded.

This paper is structured as follows: In the next section we construct the exact position representation corresponding to the MCR ({\ref{mod}, \ref{class}), and for comparison also for the Snyder/Anti-Snyder cases, and use it to solve for the free particle spectrum in Sect.(3) and the infinite well spectrum in Sect.(4). For the infinite well we find that the bound state spectrum terminates at finite energy when $\alpha>0$, just as  for the harmonic oscillator case in Ref.\cite{spectra}. The spectrum is also limited above for the Anti-Snyder case, but not for the Snyder case, as expected. Furthermore, we show that the results are exactly reproduced by a semi-classical analysis in Appendix A.

An interesting feature of Dirichlet boundary conditions for the infinite well with MCR (\ref{mod}) is that the Hamiltonian is not Hermitian though the energy eigenvalues are real! In Sect.(4.3) we show that periodic boundary conditions, which equivalently describe a particle confined in a thin ring, give rise to a Hermitian Hamiltonian. 

In Sect.(5) we discuss a puzzle we uncovered from the literature: A perturbative calculation of the energy spectrum of the infinite well for the MCR (\ref{mod}), using the expansion (\ref{lowmom}) and standard textbook formulae, gives a result \cite{Pedram2} which does not agree with the perturbative limits of the exact result from Sect.(4). We trace the discrepancy to the fact that the perturbative Hamiltonian, just like the exact Hamiltonian, is not Hermitian. A correct perturbative calculation requires the use of a more fundamental formula than the final compact expression found in textbooks.

As not all problems are easily handled using the exact position representation, in Sect.(6) we illustrate how the low-momentum expansion, when supplemented with the consistency condition (\ref{const}), can be used to obtain the correct spectrum of the infinite well for low-momentum modes. We also use the low-momentum expansion to study scattering from a step potential.

In Sect.(7) we discuss the formation of bound states in an asymmetric finite well. We show that the MCR (\ref{mod}) delays the formation of bound states compared to usual quantum mechanics, just as for the Snyder case. However for the Anti-Snyder case,  the formation of bound states is actually enhanced.   

In Sect.(8) we discuss a second puzzle from the literature: In Ref.\cite{das}, as a result of a low momentum analysis, it was stated that an MCR of the form (\ref{mod},\ref{class}) implied the quantisation of the length of an infinite well, and thus the discretisation of space. Our exact solutions of the infinite well and the particle in a ring in Sect.(4), and the harmonic oscillator in Ref.\cite{spectra}, do not display such a quantisation. Resolving this puzzle uncovers a number of intricacies, with our conclusion differing from that of Ref.\cite{das}.  
 
We summarise our main results in Sect.(9) while in Appendix B we discuss the relationship between  Hermiticity and current conservation for the deformed Hamiltonian.

\section{Position Representation}
We will focus on cases $q \le 1$ in (\ref{class}) which show maximum momentum\footnote{Since for $q \le 1$ there is no minimum position uncertainty, there are no related conceptual problems in considering sharp boundaries in the problems below.}. Substituting $X=x$ and $P \equiv F(p)$ in (\ref{mod},\ref{class}) gives the differential equation 
\begin{equation}
{d F \over dp } = \left(1 -2\alpha F+ q \alpha^2 F^2\right) \, 
\end{equation}
which is easily integrated. However as the  expression for general $q$ is cumbersome,  we exhibit below the results for only some specific $q$'s.   

For $q=0$,
\begin{eqnarray}
P= {1 \over 2 \alpha} ( 1- e^{-2\alpha p}) \, , \label{q0}
\end{eqnarray}
while for $q=1$
\begin{eqnarray}
P= {p \over 1 +  \alpha p}  \, . \label{q1}
\end{eqnarray}
as one can also verify by direct substitution in (\ref{mod}). The reader should keep in mind that both $P$ and $p$ are operators which appear in the commutation relations (\ref{postrep1}-\ref{postrep2}). If one takes the Fourier transform of $P$ in (\ref{q0}) or {\ref{q1}) then one ends up with the ``$\rho$-representation" that we discussed in Ref.\cite{spectra}.

For general $q$, by writing $P= \sum_{n=1} a_n p^n $ and substituting in (\ref{mod}) we obtain the low-momentum expansions 
\begin{eqnarray}
P &=& p \left( 1- \alpha p + {(2+q) \over 3} (\alpha p)^2 + ... \right) \, , \\
P^2 &=& p^2 \left( 1- 2\alpha p  + {(7+2q) \over 3} (\alpha p)^2 + O(\alpha p)^3) ...\right)  \, , \label{free}
\end{eqnarray}
valid for $\alpha p \ll 1$.  It is interesting to note that for  for $q=-7/2$ the Hamiltonian for a free particle, $H=P^2$, takes the form  
$H= p^2( 1-2\alpha p + O(\alpha p)^3) $ with no term of order $\alpha^2$; we will use this fact to simplify our calculations in Sect.(6.1).   

\section{Plane Waves}

As a first application of the position representation, we discuss the free particle states.
Since the momentum operator $P(p)$ commutes with $p$, we may choose its eigenstates to be the plane waves $e^{ikx}$. We assume that $P$ is the physical momentum operator and so its eigenvalues $\lambda_p$ are required to be real\footnote{We use $k$ to denote the c-number wavevector and reserve the letters $P$ and $p$ for the operators.},
\begin{eqnarray}
P e^{ikx} &=&  \lambda_p e^{ikx} \, , \label{mome}
\end{eqnarray}
which implies that $\lambda_p = {1 \over 2 \alpha} ( 1- e^{-2\alpha k})$ for $q=0$. Similarly for $q=1$, $\lambda_p = {k \over 1 + \alpha k}$.  We see that $k$ must be real if $\lambda_p$ is real. 

\subsection{Free Particle}
Since the Hamiltonian $H=P^2$ for a free particle commutes with the momentum operator $P$, the plane waves are simultaneously energy eigenstates. The eigenvalue equation $H \psi =E \psi$ gives the energy, which for $q=0$ is 
\begin{eqnarray}
E &=& \left( 1- e^{-2\alpha k} \over 2 \alpha \right)^2 \, , \label{qe0}
\end{eqnarray}
and hence the eigenstates are given by 
\begin{equation}
\psi_1(x)= e^{ik_1 x} \, \hspace{1cm} \mbox{and} \, \hspace{1cm} \psi_2(x)=e^{ik_2 x}  
\end{equation}
with
\begin{eqnarray}
k_1 &=& {-1 \over 2 \alpha} \ln( 1 + 2\alpha \sqrt{E}) \, , \label{k1} \\
k_2 &=& {-1 \over 2 \alpha} \ln( 1 - 2\alpha \sqrt{E}) \, . \label{k2}
\end{eqnarray}
Now, since plane waves are momentum eigenstates, so Eq.(2) implies \cite{spectra} that for $q=0, P_{max} =\lambda_{p(max)} <1/(2\alpha)$, which implies from Eqs.(\ref{mome}-\ref{qe0}) that $0< \sqrt{E} < 1/(2\alpha)$.  Hence in Eqs.(\ref{k1}-\ref{k2}), $-(\ln 2) /2\alpha   < k_1 < 0 \ $ and $\ 0 < k_2 < \infty $.

Similarly, for $q=1$ we get
\begin{eqnarray}
E= \left({k \over 1 +  \alpha k} \right)^2  \, , \label{qe1}
\end{eqnarray}
with the wavenumbers
\begin{eqnarray}
k_1 &=& {-\sqrt{E} \over 1 + \alpha \sqrt{E} } \, , \label{k11} \\
k_2 &=& {\sqrt{E} \over 1 - \alpha \sqrt{E} } \, . \label{k21}
\end{eqnarray}
Since for $q=1, P_{max} = 1/(\alpha)$, so (\ref{qe1}) implies $0< \sqrt{E} < 1/\alpha$ and in Eqs.(\ref{k11}-\ref{k21}) the range is $-1/(2\alpha)  < k_1 < 0 \ $ and $\ 0 < k_2 < \infty $.

Although in a series expansion the Hamiltonian $H$ is a differential operator of infinite order, it is still linear. Hence the general free particle solution is given by the superposition
\begin{equation}
\psi(x) = A e^{ik_1 x} + B e^{ik_2 x} \, .  \label{sup}
\end{equation}

We remark that the limits on energies of pure plane waves (momentum eigenstates) mentioned above will not be implemented when we use a superposition of plane wave solutions in later sections, for example for the infinite well. In those later cases the constraints on the discrete spectrum energies nonetheless arise.  

\subsection{String Motivated and Anti-Snyder MCR's}

For later comparison with the results of (\ref{class}), we discuss here the MCR's
\begin{eqnarray}
[X,P]&=& i \left(1 \pm \beta P^2\right)\label{sas}
\end{eqnarray} 
with $\beta >0$. The upper sign corresponds to the string-motivated case studied in \cite{kempf,snyder} while the lower sign is the Anti-Snyder case of Ref.\cite{mig} which exhibits a maximum momentum. 

The momentum operator in the coordinate representation $X=x$ for the two cases is given by
\begin{eqnarray}
P &=& { \tan(\sqrt{\beta} p) \over \sqrt{\beta} } \,  \hspace{1cm}  \mbox{(Snyder)} \, , \nonumber \\
P &=& { \tanh(\sqrt{\beta} p) \over \sqrt{\beta} }\,  \hspace{1cm}  \mbox{(Anti-Snyder)} \nonumber \, .
\end{eqnarray}

The free-particle wavevectors are
\begin{eqnarray}
k_2 &=& -k_1 = { \tan^{-1}(\sqrt{\beta E}) \over \sqrt{\beta} }\,  \hspace{1cm}  \mbox{(Snyder)} \, , \nonumber \\
k_2 &=& -k_1 =  { \tanh^{-1}(\sqrt{\beta E }) \over \sqrt{\beta} }\, \hspace{1cm}  \mbox{(Anti-Snyder)} \nonumber \, .
\end{eqnarray}

It is noted that one may obtain the Anti-Snyder results from the Snyder case by the replacement $\sqrt{\beta} \to i  \sqrt{\beta}$.

\section{Particle in a box}
In this section we use the position representation to solve exactly for the spectrum of  the infinite well with walls 
at $x=0,L$. Using (\ref{sup}) and imposing the usual vanishing of the wavefunction at the boundaries gives the quantisation condition
\begin{equation}
k_2 -k_1 = {2 n \pi \over L} \, , \label{quant}
\end{equation}
with $n$ an integer and the corresponding eigenfunction
\begin{equation}
\psi_n(x) = {-i \over \sqrt{2L}} (  e^{ik_2 x} -  e^{ik_1 x} ) \, .  \label{iwell}
\end{equation}

For $q=0$ the above expressions, and those form the last section, imply the exact energy eigenvalues
\begin{equation}
E_n = \left( { 1 \over 2 \alpha} \right)^2 \tanh^2 \left({2 n \pi \alpha \over L} \right) \, . \label{q0box}
\end{equation}

In Appendix A we show that a semi-classical evaluation gives the same result as (\ref{q0box}). A perturbative expansion of  (\ref{q0box}) gives
\begin{equation}
E_n = \left( { n \pi  \over L} \right)^2  -  {8\alpha^2 \over 3} \left({ n \pi \over L}\right)^4  + O(\alpha )^4 \, . \label{q0boxpert}
\end{equation}
 
Similarly for $q=1$ the exact eigenvalues are 
\begin{equation}
E_n = \left({ -1 + \sqrt{ 1 +  ({2 \alpha n \pi \over L})^2 } \over 2 \alpha^2 ({n \pi \over L}) } \right)^2    \label{q1box}
\end{equation}
which also agree with the semi-classical expression in Appendix A.

Notice that as $n \to \infty $, the energies in (\ref{q0box}, \ref{q1box}) attain finite values. Thus unlike usual quantum mechanics, the bound state energies are limited above, just as we found for the harmonic oscillator in \cite{spectra}. 

Recall that in the {\it deformed} quantum theory with MCR (\ref{mod}) the Hamiltonian is a differential operator of infinite order. One may understand the severe departure of the spectrum from usual quantum theory by re-writing $H=p^2 + (P^2 - p^2) $ and interpret this as the Hamiltonian of usual quantum mechanics but with a momentum dependent potential $V(P)=P^2-p^2$; the momentum dependence is not small (perturbative) for large momentum states, thus causing a strong deformation of the spectrum.

By comparison, for the Snyder/Anti-Snyder cases it is easily shown that the exact bound state energies in infinite well are
\begin{eqnarray}
E_n &=& { \tan^2({\sqrt{\beta} n \pi \over L})  \over \beta} \, \hspace{1cm}  \mbox{(Snyder)} \, ,\nonumber \\ 
E_n &=& { \tanh^2({\sqrt{\beta} n \pi \over L})  \over \beta} \, \hspace{1cm}  \mbox{(Anti-Snyder)} \, , \nonumber
\end{eqnarray}
which also agree perfectly with the corresponding semi-classical results. The leading order terms in a $\beta$-expansion of these expressions also  agree with standard perturbation theory as one may verify. Notice that the energies are limited above for the Anti-Snyder case but not for the Snyder case. 

\subsection{Expectation values and uncertainties}
For $q=1$, the position expectation values in an eigenstate are
\begin{eqnarray}
\langle X \rangle&=&\frac{1}{L}\int_{0}^{L}dx\ x \left[1-\cos\Bigl(\frac{2 n\pi x}{L}\Bigr)\right]=\frac{L}{2} \nonumber \, , \\
\langle X^2  \rangle&=&\frac{1}{L}\int_{0}^{L}dx\ x^2 \left[1-\cos\Bigl(\frac{2 n\pi x}{L}\Bigr)\right]=\Bigl(\frac{1}{3}-\frac{1}{2n^2\pi^2}\Bigr)L^2   \nonumber \, , \\
\Rightarrow \Delta X&=&\sqrt{\langle X^2\rangle-\langle X \rangle^2}=\frac{L}{2}\sqrt{\frac{1}{3}-\frac{2}{n^2\pi^2}} \nonumber \, , 
\end{eqnarray}
which are the same as in the undeformed, $\alpha=0$ case. The uncertainty in position increases with $n$ and saturates at $\bigl(\Delta X\bigr)_{n\rightarrow \infty}=L/(2\sqrt{3})$. 

For the momentum operators we find after some algebra
\begin{eqnarray}
\langle P \rangle&=& 0 \, , \nonumber \\
\langle P^2 \rangle&=&   \langle H \rangle  \nonumber \\
&=& E_n \nonumber \\
&=& \left[\frac{-1+\sqrt{1+4\alpha^2\bigl(n^2\pi^2/L^2\bigr)}}{2\alpha^2\bigl(n\pi/L\bigr)}\right]^2 \, , \nonumber\\
\Rightarrow \Delta P&=&\sqrt{\frac{1}{\alpha^2}+\frac{L^2\bigl(1-\sqrt{1+\frac{4n^2\pi^2\alpha^2}{L}}\bigr)}{2n^2\pi^2\alpha^4}} \nonumber \, .
\end{eqnarray}  
As for the uncertainty in position, $\Delta P$ increases monotonically with $n$, reaching it's maximum at $(\Delta P)_{n\rightarrow\infty}=1/\alpha$. Thus the product of the uncertainty of position and momentum is thus given by
\begin{eqnarray}
(\Delta X)(\Delta P)&=&\frac{L}{2}\sqrt{\left(\frac{1}{\alpha^2}+\frac{L^2\bigl(1-\sqrt{1+\frac{4n^2\pi^2\alpha^2}{L}}\bigr)}{2n^2\pi^2\alpha^4}\right)\times\left(\frac{1}{3}-\frac{2}{n^2\pi^2}\right)}.
\end{eqnarray}

The vanishing of $\langle P \rangle$ for the infinite well is different from the situation encountered for the harmonic oscillator \cite{spectra}. This leads to other differences: For the $q=1$ 
harmonic oscillator the  uncertainties vanished at the top of the bound state spectrum, corresponding to $P=P_{max}$ and the vanishing of the MCR, suggesting classical characteristics. However we see that for the $q=1$ infinite well the uncertainties do not vanish at the top of the bound state spectrum.

For comparison, let us consider now the Anti-Snyder case. The values for $\langle X \rangle$ and $\langle X^2 \rangle$
are the same as for undeformed quantum mechanics and hence the same as the $q=1$ case above.  For the momentum operator
\begin{eqnarray}
\langle P \rangle &=& 0 \nonumber \, , \\
\Rightarrow \Delta P&=&\frac{\tanh\Bigl(\frac{\sqrt{\beta}n\pi}{L}\Bigr)}{\sqrt{\beta}} \nonumber \, .
\end{eqnarray}  
$\Delta P$ increases with states $n$ and reaches it's maximum $(\Delta P)_{n\rightarrow\infty}=1/\sqrt{\beta}$. The product of the uncertainty of position and momentum is given by
\begin{eqnarray}
(\Delta X)(\Delta P)&=&\frac{\tanh\Bigl(\frac{\sqrt{\beta}n\pi}{L}\Bigr)L}{2\sqrt{\beta}}\sqrt{\frac{1}{3}-\frac{2}{n^2\pi^2}}.
\end{eqnarray}
It too does not vanish at the top of the bound state spectrum.

\subsection{Orthogonality}

It is easy to verify that for $\alpha \neq 0$, the eigenstates for different $n$ (\ref{iwell}) corresponding to the infinite well problem above are {\it not} orthogonal! This is related to the fact that though the eigenvalues of $H=P^2$ are real, $H$ is not Hermitian\footnote{We are using common physics terminology here. More precisely, the Hamiltonian is not ``symmetric"; a stronger condition of self-adjointness is also often discussed, see \cite{kempf,Pedramx}.} on the space of eigenstates: That is, except for $m =n$, 
\begin{equation}
\langle n| H| m\rangle \neq   \langle m| H| n\rangle^{*} \, . \label{herm}
\end{equation}
 We will see this more explicitly in Sect.(5).

Though the Dirichlet boundary conditions lead to a non-Hermitian Hamiltonian for the MCR (\ref{mod}), the boundary conditions may be motivated thorough a physical limiting procedure as follows: For a general finite smooth potential $V$ the Schrodinger equation implies 
\begin{equation}
\psi = { (E-P^2)\psi \over V}  \, . 
\end{equation}
For finite $E$ and as $V \to \infty$, we find $\psi \to 0$ as long as $P^2 \psi$ remains finite. Furthermore, as one has real energies which interpolate smoothly to the $\alpha \to 0$ case despite the non-Hermiticity, we feel the Dirichlet boundary conditions are acceptable. Nevertheless, we discuss alternative Hermiticity-preserving boundary conditions in the next sub-section.

For the Snyder/Anti-Snyder case, since $k_2 = -k_1$ the eigenstates (\ref{iwell}) are orthogonal and the Hamiltonian is Hermitian.

 \subsection{Periodic Boundary Conditions}
 
 A particle in a one-dimensional box with periodic boundary conditions is equivalent to a particle confined to a ring of infinitesimal thickness. Imposing on the wavefunction (\ref{sup}) and its derivative the usual single-valuedness condition as one traverses the ring gives the familiar expressions
\begin{equation}
\psi_n(x) = { e^{ik^{(n)} x}  \over \sqrt{2\pi R} } \, ,  \label{pbc}
\end{equation}
with 
\begin{eqnarray}
k^{(n)} &=& { n \over R} \, , \label{pbck}
\end{eqnarray}
and $n$ an integer. We have taken the length of the box to be $L=2\pi R$.  For $q=1$, the existence of $P_{max}$ for momentum eigenstates, as for free plane waves, restricts $k$ and hence we have $-R/(2\alpha) < n < \infty $. The corresponding energies follow from ({\ref{qe1}): 
\begin{equation}
E_n = \left({n \over R}\right)^2 \left( 1 + {\alpha n \over R}\right)^{-2}\, , \hspace{1cm}  -R/(2\alpha) < n < \infty \, .
\end{equation} 
The spectrum is again bounded above as $n \to \infty $. 

Notice that unlike the Dirichlet boundary conditions for the infinite well, the periodic conditions result in a single plane wave of the form (\ref{pbc}) with the usual quantised wavevectors (\ref{pbck}). Thus in this case the eigenstates are orthonormal and the Hamiltonian is Hermitian; the two sides of ({\ref{herm}) are now equal. It is also worth noting for later comparison that the solution ({\ref{pbck}) is infinitely differentiable as required for an exact solution of a differential operator (the Hamiltonian) of infinite order.

\subsubsection{Uncertainties}

For the particle in a ring (\ref{pbc}), the position expectation values in an eigenstate $n$ are independent of $n$ and are the same as for usual $\alpha=\beta =0$ quantum mechanics:  
\begin{eqnarray}
\langle X\rangle&=& \pi R    \ , \nonumber\\
\langle X^2\rangle&=& \frac {4} {3} (\pi R)^2 \ , \nonumber\\
\Delta X &=& \sqrt{ \langle X^2\rangle-  \langle X\rangle^2 } =   \frac {\pi R} {\sqrt{3}} \ . \nonumber
\end{eqnarray}
Since the $n$ states are momentum eigenstates, $\langle P^2 \rangle = \langle P \rangle^2 = \lambda_{p}^2$ as defined in Eq.(\ref{mome}). So $\Delta P \equiv 0$.  These statements apply also to the Snyder/Anti-Snyder algebras.

\section{First Puzzle: Failure of Perturbation Theory?}

In usual $\alpha=0$ quantum mechanics the infinite well eigenstates are $\langle x|n \rangle = \psi_{n(0)} = \sqrt{2 \over L} \sin({n \pi x \over L})$ with energies $E_{n(0)} = (n \pi /L)^2 $. One may treat the deformation (\ref{mod}) perturbatively and so obtain the energy shifts. 

At leading order, standard perturbation theory \cite{messiah} gives the perturbed wavefunction 
\begin{eqnarray}
\psi_{n(1)} &=& \psi_{n(0)} + \sum_{m \neq n} a_m \psi_{m(0)}  \, \label{pertwave} 
\end{eqnarray}
with
\begin{eqnarray}
a_m &=& { \langle m | H_1 | n \rangle \over   E_{n(0)} - E_{m(0)} }  \, , \hspace{1cm}   m\neq n \, .
\end{eqnarray}
From (\ref{free}), $H=H_0 + H_1 + H_2 +...$. With $H_1 = -2\alpha p^3 = -2\alpha (-i \partial_x)^3$ one gets
\begin{eqnarray}
 \langle m | H_1 | n \rangle &=& { i8\alpha \pi^2 \over L^3} { mn^3 \over m^2-n^2} \, , \label{h1} \\
 a_{m}^{pert} &=& { -i 8 \alpha \over L} { mn^3 \over (m^2-n^2)^2 } \,  \hspace{0.5cm}  \mbox{for} \; m-n = \mbox{odd}  \label{apert}
\end{eqnarray} 
and zero otherwise. As a check one may also determine $a_m$, as defined by (\ref{pertwave}), by projecting the exact solution (\ref{iwell}) on an unperturbed state and taking the leading term as $\alpha \to 0$,
\begin{equation}
a_m = \int_0^L \psi_{m(0)} \psi_{n}^{exact} |_{\alpha \to 0} \label{check} \, . 
\end{equation}
For $q=1$ we have verified that (\ref{check}) gives the same expression as (\ref{apert}). 

Since $\langle n|H|n\rangle =0$, up to order $\alpha^2$ the perturbed energies are given by 
\begin{eqnarray}
E_n &=& E_{n(0)} + E_{n}^{(2,1)} + E_{n}^{(1,2)} \, ,\\
E_{n}^{(2,1)} &=& \langle n | H_2 | n \rangle \, , \\
E_{n}^{(1,2)} &=& \sum_{m \neq n} { \langle m | H_1 | n \rangle \langle n | H_1 | m \rangle \over   E_{n(0)} - E_{m(0)} } \, .\label{fund}
\end{eqnarray}
If $H_1$ is Hermitian, that is, if 
\begin{equation}
\langle m | H_1 | n \rangle = \langle n | H_1 | m \rangle^{*} \label{herm1}
\end{equation}
then one obtains the usual compact textbook formula 
\begin{eqnarray}
E_{n}^{(1,2) \ (tb)} &=& \sum_{m \neq n} { | \langle n | H_1 | m \rangle |^2 \over   E_{n(0)} - E_{m(0)} }  \, . \label{tb}
\end{eqnarray}

However it is easily checked that (\ref{herm1}) is {\it not} true for the infinite well with usual Dirichlet boundary conditions: Integration by parts of the left-side  to obtain the right-side gives also surface terms that do not vanish. More simply, one can see directly from (\ref{h1}) that (\ref{herm1}) is not valid in this case. 
Indeed, if one uses the textbook formula (\ref{tb}), it  gives a result \cite{Pedram2} which does not agree with the perturbative expansion of the exact expressions of the last section. However if one uses the more fundamental expression (\ref{fund}), which is still valid, then one obtains
\begin{eqnarray}
E_{n}^{(1,2)} &=&  { -5 \alpha^2 \pi^4 n^4  \over  L^4} 
\end{eqnarray}
and since
\begin{eqnarray}
E_{n}^{(2,1)} &=&  { (7+ 2q) \alpha^2 \pi^4 n^4  \over  3 L^4} 
\end{eqnarray}
then the net result for $\Delta E= E_{n}^{(2,1)} + E_{n}^{(1,2)}$ is 
\begin{eqnarray}
\Delta E_{n}^{pert} &=&  { 2 (q-4) \alpha^2 \pi^4 n^4  \over  3 L^4} + O(\alpha^4) \, .
\end{eqnarray}
This agrees with the expansion of the exact results from previous sections and also the semi-classical results in Appendix A.
Notice the $(q-4)$ factor which was found to be a universal trend in \cite{spectra}. For $q=-7/2$ the result also agrees with the low-momentum solution of the infinite well to be discussed in the following section.

\section{Applications of the Low-Momentum Expansion}

\subsection{Infinite Well}

Here we discuss the approximate solution of the infinite well problem with Dirichlet boundary conditions using the low-momentum expansion (\ref{free}). For brevity we discuss the case with $q=-7/2$ as fewer terms need to be considered to obtain the spectrum at order $\alpha^2$. 

The corresponding (higher-order) Schroedinger equation for $q=-7/2$ is  
\begin{eqnarray}
\left(\frac{d^2}{dx^2}+ \kappa^2+ 2i\alpha\frac{d^3}{dx^3}\right)\Psi(x)&=&0 \label{6}
\end{eqnarray}
\noindent where we have defined $\kappa:=\sqrt{(E-V(x))}$ and the energy is $E= \kappa^2+V(x)$. The ignored error terms in the above equation are of order $\alpha^3$. 

For the infinite well with boundaries at $x=0$ and $x=L$, the three solutions of (\ref{6}) are of the form $\Psi(x)=e^{kx}$, with  \begin{eqnarray}
k&=& \{ k_1, k_2, k_3 \} \label{basis} \\
k_1& \equiv & -\kappa (1- \alpha \kappa + 5\alpha^2\kappa^2 /2) \, , \label{k1low} \\
k_2& \equiv & \kappa (1 + \alpha \kappa + 5\alpha^2\kappa^2 /2) \, , \label{k2low} \\
k_3& \equiv &  {1 \over 2\alpha} + 2i \alpha \kappa^2  \label{k3low} \, . 
\end{eqnarray}
Since $k_3$ does not satisfy the low-momentum consistency condition (\ref{const}), it must be discarded (in Sect.(8) we discuss the consequences of keeping $k_3$). 

The general wavefunction can thus be expressed as
\begin{eqnarray}
\Psi(x)&=& A e^{ik_{1} x}+ Be^{ik_{2}x} \label{wf1}
\end{eqnarray}  
with the wavevectors (\ref{k1low},\ref{k2low}) being smooth deformations of the usual solutions of the undeformed Schrodinger equation.

By imposing Dirichlet boundary conditions on (\ref{wf1}) the quantised energies are obtained,
\begin{eqnarray}
E_{n}&=& \frac{n^2\pi^2}{L^2}\left(1-\frac{5n^2\pi^2\alpha^2}{L^2}+O(\alpha^4)\right)\label{ExactE1}
\end{eqnarray}
with corresponding eigenfunctions 
\begin{eqnarray} 
\Psi(x)&=&\sqrt{\frac{2}{L}}\sin\Bigl(\frac{n \pi x}{L}\Bigr) \exp\left[i\Bigl( \frac{n^2\pi^2\alpha x}{L^2}+\frac{\pi}{2}\Bigr)\right]\label{wf3} \, .
\end{eqnarray}  
The energy shifts are given by
\begin{eqnarray}
\Delta E_{n}&=&-\frac{5n^4\pi^4}{L^4}\alpha^2+O(\alpha^4) \, , \label{ShiftE1}
\end{eqnarray} 
in agreement with perturbation theory of the last section and the semi-classical results of the Appendix.

\subsection{Scattering from a Potential Step}

For a second illustration of the low-momentum approach, consider a one dimensional potential step defined by $V(x)=V_{0} \Theta(x)$, where $\Theta(x)$ is the step function. Consider first the case when the particle energy $E< V_{0}$. Then for $q=-7/2$ the wavefunction to the left and the right of the barrier are determined by 
\begin{eqnarray}
(\frac{d^2}{dx^2}+  2i\alpha\frac{d^3}{dx^3} +k_{1}^2)\Psi_{<}(x)&=&0\\
(\frac{d^2}{dx^2}+ 2i\alpha\frac{d^3}{dx^3} -k_{2}^2)\Psi_{>}(x)&=&0
\end{eqnarray}
where $k_{1}=\sqrt{E}; \ $  $k_{2}=\sqrt{(V_{0}-E)}$. As in the previous subsection, the ignored error terms in the differential equations are $O(\alpha^3)$. The ansatz $\Psi=e^{mx}$ produces the algebraic equations,
\begin{eqnarray}
(m^2+2i\alpha m^3+k_{1}^2)&=&0\\
(m^2+2i\alpha m^3-k_{2}^2)&=&0
\end{eqnarray}
with the corresponding approximate solutions for the deformed wave numbers,
\begin{eqnarray}
x<0: m&=&\{ik_{1}^{'},-ik_{1}^{''},1/(2\alpha)-2k_{1}^2\alpha\}\nonumber\\
x>0: m&=&\{k_{2}^{'},-k_{2}^{''},i/(2\alpha)+2ik_{2}^2\alpha\}\label{rootm1}
\end{eqnarray}
where 
\begin{eqnarray}
k_{1}^{'}&=&k_{1}(1+k_{1}\alpha),\ k_{1}^{''}=k_{1}(1-k_{1}\alpha) \, , \nonumber\\
k_{2}^{'}&=&k_{2}(1-ik_{2}\alpha),\ k_{2}^{''}=k_{2}(1+ik_{2}\alpha) \, , 
\end{eqnarray}
to the indicated order in $\alpha$. As before, a consistent low-momentum expansion requires that we discard the third root in \eqref{rootm1} (contrast this with Ref.\cite{das}). Thus assuming the particle is approaching the barrier from the left, the  wave function is
\begin{eqnarray}
\Psi_{<}(x)&=& A e^{ik_{1}^{'}x}+B^{-ik_{1}^{''}x} \, , \\
\Psi_{>}(x)&=& C e^{-k_{2}^{''}x} \, . 
\end{eqnarray}
Continuity of wave function at $x=0$ implies $A+B=C$. 

From \eqref{current1} the conserved current is given by
\begin{eqnarray}
J_{<}&=&2k_{1} \Bigl(|A|^2-|B|^2\Bigr)- 4\alpha k_{1}^2 \Bigl(|A|^2+|B|^2\Bigr)+O(\alpha^2)\label{Current2}\\
J_{>}&=&-6\alpha^3 k_{2}^4 |C|^2=O(\alpha^3) \approx 0 .\label{Current3}
\end{eqnarray}
Continuity of probability current density at the boundary implies, to leading order  in  $\alpha$, that
\begin{eqnarray}
\frac{|B|^2}{|A|^2}&=&\frac{1-2\alpha k_{1}}{1+2\alpha k_{1}}.\label{1a}
\end{eqnarray}  

On the left of the potential step, the incident and reflected current densities are given by,
\begin{eqnarray}
J_{\text{inc}}&:=& 2k_{1} \bigl(1-2\alpha k_{1}\bigr)|A|^2 \, , \nonumber\\
J_{\text{ref}}&:=& 2k_{1} \bigl(1+2\alpha k_{1}\bigr)|B|^2 \, ,
\end{eqnarray} 
while in the region $x>0$ the transmitted current density is
\begin{eqnarray}
J_{\text{trans}}&:=&O(\alpha^3)= 0.
\end{eqnarray} 
Thus the reflection coefficient is given by 
\begin{eqnarray}
R&:=&\left|\frac{J_{\text{ref}}}{J_{\text{inc}}}\right|=\frac{\bigl(1+2\alpha k_{1}\bigr)|B|^2}{\bigl(1-2\alpha k_{1}\bigr)|A|^2}\nonumber\\
&=&1 \label{R1}
\end{eqnarray}
and since the transmission coefficient $T$ vanishes at the order calculated, we have $R+T=1$.
Hence for $q=-7/2$ the modified commutation relation does not affect the $R$ and $T$ coefficients to the order calculated. 

Next consider the case when the energy of the particle is above the potential barrier, $E>V_{0}$. Now the wavefunction is  
\begin{eqnarray}
\Psi_{<}(x)&=& A e^{ik_{1}^{'}x}+B^{-ik_{1}^{''}x};\\
\Psi_{>}(x)&=& C e^{-ik_{2}^{''}x}
\end{eqnarray}
with 
\begin{eqnarray}
k_{1}^{'}&=&k_{1}(1+k_{1}\alpha);\  k_{1}^{''}=k_{1}(1-k_{1}\alpha)\nonumber\\
k_{2}^{''}&=&k_{2}(1+k_{2}\alpha)\nonumber\\
k_{1}&=&\sqrt{E};\ k_{2}=\sqrt{(E-V_{0})}.
\end{eqnarray}
The continuity of the wave function and conserved current give,
\begin{eqnarray}
A+B&=&C\\ 
k_{1} \Bigl(|A|^2-|B|^2\Bigr)- 2\alpha k_{1}^2 \Bigl(|A|^2+|B|^2\Bigr)&=& k_{2}|C|^2\Bigl(1-2\alpha k_{2}\Bigr),
\end{eqnarray}
and thus we obtain the reflection and transmission coefficients 
\begin{eqnarray}
R&=&\frac{(k_{1}-k_{2})^2}{(k_{1}+k_{2})^2}\Bigl(1-12k_{1}k_{2}\alpha^2 \Bigr)\\
T&=&\frac{4k_{1}k_{2}}{(k_{1}+k_{2})^2}\Bigl(1+3(k_{1}-k_{2})^2\alpha^2 \Bigr)\\
\Rightarrow R+T&=&1.
\end{eqnarray}

\section{Finite Potential Well}

We now study the occurrence of bound states inside an asymmetric potential well defined by $V(0<x<L)=-V_0, \, V(x>L)= 0$ and $V(0) = \infty$ with $V_0 >0$. So the bound state energies will be $E <0$.
For region I, where $0<x <L$, the wavefunction will be of the form (\ref{sup}) and demanding that it vanishes at $x=0$ gives
\begin{equation}
\psi(x)_I = A (e^{ik_2 x} - e^{ik_1 x} ) \, .  \label{fwell-1}
\end{equation}
For $q=1$, 
\begin{eqnarray}
k_1 &=& {-\sqrt{E+V_0} \over 1 + \alpha \sqrt{E+V_0} } \, , \label{k1w} \\
k_2 &=& {\sqrt{E+V_0} \over 1 - \alpha \sqrt{E+V_0} } \, , \label{k2w}
\end{eqnarray}
Note $E+ V_0 >0$. In region II, $x>L$, the wavefunction is 
 \begin{equation}
\psi(x)_{II} = C e^{ik_3 x}  \, ,  \label{fwell-2}
\end{equation}
with $Im[k_3] >0$ required for a bound (normalisable) state. For $q=1$,
\begin{eqnarray}
k_3 &=& {i\sqrt{-E} \over 1 - \alpha \sqrt{-E} } \, . \label{k3w} 
\end{eqnarray}

All the expressions above reduce to the familiar ones when $\alpha=0$. Matching the wavefunction  and its slope at $x=L$ gives two conditions which we write below for the most general case, allowing the wavevectors to be complex (that is allowing $E+V_0 <0$). Defining $\Delta k \equiv k_2 -k_1$:
\begin{eqnarray}
Re[k_1 + k_2 ] + Im \large[ \Delta k \cot \left({\Delta k L \over2}\right) \large] &=& 2 Re[k_3] \label{c1} \\
Im[k_1 + k_2 ] - Re \large[ \Delta k \cot \left({\Delta k L \over2} \right) \large]&=& 2 Im[k_3] \label{c2} 
\end{eqnarray}
From (\ref{k1w},\ref{k2w}) we see that $k_1 + k_2$ is always real. For $E+V_0 <0$, $\Delta k$ is purely imaginary and then one can easily show graphically  that condition (\ref{c2}) has no solution for bound states, $Im[k_3] >0$. On the other hand, for the usual case  $E+V_0 >0$, condition (\ref{c2}) implies 
\begin{equation}
\tan \left({\Delta k L \over2}\right) = { -\Delta k \over 2 Im[k_3]}
\end{equation} 
which is seen graphically to require
\begin{equation}
\Delta k > {\pi \over L} \, ,   \label{bound}
\end{equation}
a condition identical to that for the $\alpha=0$ theory. However we must also satisfy condition (\ref{c1}) which for $q=1$ is 
\begin{equation}
{ 2 \alpha (E+V_0) \over 1-\alpha^2(E+V_0) } = { 2 \alpha E \over 1-\alpha^2 E }
\end{equation}
which is an identity for $\alpha=0$ but has no viable solution for $\alpha \neq 0$. Hence for $\alpha \neq 0$ bound state formation is inhibited for $q=1$. A similar conclusion follows for $q=0$ and we believe this is a general property of (\ref{mod}) for $\alpha \neq 0$.  

However, although in the above analysis we used exact expressions for the wavefunctions on both sides of the boundary, the sharp boundary makes it impossible to match all derivatives of the wavefunction at the boundary. So it is probably more sensible to interpret the above results as holding in a low-momentum approximation where only a few derivatives are matched.  We expect that for a smooth boundary, bound states in an asymmetric well for the MCR (\ref{mod}) will still be permitted though possibly ``delayed" as discussed below for the Snyder case.

\subsection{Snyder/Anti-Snyder Case}
 
The relevant wavevectors for the same asymmetric finite well are now
 \begin{eqnarray}
k_2 &=& -k_1 = { \tan^{-1}(\sqrt{\beta (E+V_0) }) \over \sqrt{\beta} }\,  \hspace{1cm}  \mbox{(Snyder)} \, , \\
k_1 &=& -k_1 =  { \tanh^{-1}(\sqrt{\beta (E+V_0)}) \over \sqrt{\beta} }\,  \hspace{1cm}  \mbox{(Anti-Snyder)} \, 
\end{eqnarray}
and 
\begin{eqnarray}
k_3 &=& { \tan^{-1}(i\sqrt{-\beta E }) \over \sqrt{\beta} }\,  \hspace{1cm}  \mbox{(Snyder)} \, , \\
k_3 &=& { \tanh^{-1}(\sqrt{-\beta E}) \over \sqrt{\beta} }\,  \hspace{1cm}  \mbox{(Anti-Snyder)} \, .
\end{eqnarray}

We see that with $E+ V_0 >0$ condition (\ref{c1}) is identically satisfied while (\ref{c2}) gives the same constraint as (\ref{bound}) which for the Snyder case
implies
\begin{equation}
V_0 > { \tan^2({\alpha \pi \over 2L}) \over \alpha^2}  \, .
\end{equation}
Since for $\alpha>0$ the right-hand-side of the inequality is larger than the $\alpha=0$ limit $\pi^2/(2L)^2$, this means that formation of bound states in this finite well is delayed in the Snyder case compared to normal quantum mechanics. 

For the Anti-Snyder case the inequality (\ref{bound}) translates to   
\begin{equation}
V_0 > { \tanh^2({\alpha \pi \over 2L}) \over \alpha^2}  \, .
\end{equation}
Now we see that bound state formation is actually enhanced compared to the $\alpha =0$ case.

\section{Second Puzzle: Discretisation of space?}

Let us first summarise Ref.\cite{das}: The authors solved for the infinite well with the MCR (\ref{mod},\ref{class}) in the low-momentum approximation, but unlike our procedure of Sect.(6), 
in Ref.\cite{das} the $k_3$ solution was not discarded\footnote{In Ref.\cite{das} they solved the problem for $q=4$ but since their main conclusion comes from the $O(\alpha)$ piece of (\ref{free}), which is $q-$independent, this does not qualitatively affect the following discussion.}.

So instead of (\ref{wf1}), in Ref.\cite{das} the general solution inside the well is of the form $Ae^{ik_1 x} + B e^{ik_2 x} + C e^{ik_3 x}$, requiring the additional constant $C \equiv |C| e^{i\theta}$. Since the only boundary condition imposed in Ref.\cite{das} was the vanishing of the wavefunction at the walls, this was not sufficient to determine all the constants. The authors then assumed that $|C| \to 0$ as $\alpha \to 0$ and also that $\kappa L = n\pi + \epsilon$ 
with $\epsilon \to 0$ as $\alpha \to 0$. Following those assumptions, the authors obtained a quantisation condition which we write as
\begin{eqnarray}
k_3 L = r \pi +  O(\alpha) \, , \label{kq}
\end{eqnarray}
where $r$ is an integer. Since $k_3$ is given by (\ref{k3low}), the quantisation condition may also be written as
\begin{equation}
{L \over 2\alpha} = r \pi + O(\alpha) \, , \label{kq2}
\end{equation}
which is the form used in Ref.\cite{das}, and which they interpreted as implying a quantisation of the length of the well and hence a discreteness of space. 

We now comment on the analysis and conclusion of Ref.\cite{das}. Simply put, by keeping $k_3$, in our opinion the authors of Ref.\cite{das} were not adopting a consistent low-momentum approximation as we have discussed in the earlier sections of this paper. 
As we showed in Sect.(6.1), a consistent low-momentum calculation for the infinite well agrees with perturbative and semi-classical calculations.

But what if one treated the third-order linear differential equation (\ref{6}) as the starting point and ignored (\ref{class})? Then would not  the deformed Schrodinger equation have three independent solutions which should all be considered when constructing the general solution? However note that at leading order, which concerns us here, $k_3= 1/2\alpha$ is not an independent solution but is rather just a special case of the wavevector $k_2$.

This means that (\ref{basis}) is, to leading order, an over-complete basis. All of this is related to the fact that the Hamiltonian for the infinite well with Dirichlet boundary conditions is not Hermitian, as we discussed earlier. Nevertheless, it is in principle possible to work with an over-complete basis as long as one imposes sufficient consistency conditions. In the present situation, if all solutions of the third-order Schrodinger equation are kept (though this is inconsistent from our point of view), then one should also consider seriously the deformed current (\ref{current1}) which requires additional conditions for it to vanish at the boundaries.

One can show, through some tedious but straightforward algebra,  that if the deformed current (\ref{current1}) is required to vanish at each of the walls, and without using the external assumptions on $C,\kappa$ made in Ref.\cite{das}, then all the constants are determined in terms of the free parameter $\theta$.  If further one requires that the deformation (\ref{mod}) is smooth, then using either perturbation theory or the semi-classical expansion fixes $\theta = O(\alpha)$. Hence in some sense, even if $k_3$ is kept (though it is redundant and inconsistent to do so), one can fix the larger number of undetermined constants by imposing sufficient consistency conditions (of course the whole procedure is not very useful as far as determining the spectrum is concerned). 

What about the quantisation of space suggested in Ref.\cite{das} through (\ref{kq2})? As we have indicated by our Eq.(\ref{kq}), the condition (\ref{kq2})  is simply the usual quantisation condition on wavevectors: Recall that $k_3$ is just one specific momentum mode of $k_2$. In other words, by keeping the spurious $k_3$ mode, {\it and} writing is as in (\ref{kq2}) the authors of Ref.\cite{das} were led, in our opinion, to an erroneous conclusion (and this error is the same when considering higher orders \cite{das}). 

In summary, our solution of the infinite well problem with MCR (\ref{mod},\ref{class}), whether exactly or in a consistent low-momentum approximation, either with Dirichlet or periodic boundary conditions, does not exhibit any quantisation of length, contrary to the statement in Ref.\cite{das}.  Neither did we find any quantisation of length in an exact solution of the harmonic oscillator in Ref.\cite{spectra}.

\section{Conclusion}

In our previous paper \cite{spectra} we investigated the MCR (\ref{mod},\ref{class}) in the momentum representation and found a number of interesting results: For $q \le 1$  the bound state spectrum of concave potentials terminated at finite energy, unlike the case of usual quantum mechanics. In particular, we solved exactly for the harmonic oscillator spectrum and found that the position and momentum uncertainties vanished at the top of the bound state spectrum.

In Ref.\cite{spectra} we also calculated the semi-classical energy shifts due to the deformation and showed that for certain ranges of $q$ the energy shifts were of opposite sign to the string-motivated (Snyder) MCR, suggesting a potential means of empirically differentiating a sub-class of such deformed theories.

In this paper we studied a number of different problems corresponding to the same MCR (\ref{class}) but in the position representation. For example, we obtained the exact bound state spectrum for the infinite well; while that spectrum also terminated at finite energy, the corresponding position and momentum uncertainties did not.

We also found that the deformation (\ref{class}) delays the formation of bound states in asymmetric finite wells, just as for the Snyder (string) case. However, interestingly, for the Anti-Snyder case, we found bound state formation to be enhanced. This fact might potentially be empirically useful.

Since exact solutions are not always feasible, in Sect.(6) we discussed solutions of two problems using consistent low-momentum expansions of the MCR (\ref{class}). Consistency means that solutions of the corresponding deformed Schrodinger equation must  be restricted by the relevant low-momentum constraint. We showed that such an analysis produces results for the infinite well which agree with appropriate limits of the exact or semi-classical results.

We also examined two puzzles that we came across in the literature. The first concerned a perturbative calculation of the infinite well energy spectrum for the deformation (\ref{class}). We found that the compact textbook formula for second-order perturbation theory is inapplicable for that problem because the corresponding Hamiltonian is not Hermitian when the usual Dirichlet boundary conditions are used. Nevertheless perturbation theory is valid if one uses a more fundamental formula.  

The second puzzle we examined was a conclusion of Ref.\cite{das} which stated that the infinite well problem for the MCR (\ref{class}) implied a quantisation of the length of the well. However, through our detailed investigation of the infinite well using exact, semiclassical and consistent low-momentum methods, we did not find any evidence of length quantisation; instead we found several effects of the intrinsic maximum momentum when $q \le 1$. 

We examined the analysis of Ref.\cite{das} in Sect.(8): The authors did not implement the required low-momentum constraint (\ref{const}) on their solutions, and we believe they mis-interpreted Eq.(\ref{kq}).

As a result of our conclusions concerning the two puzzles mentioned above, we feel that other related calculations in the literature \cite{das, das-rel} probably deserve a careful re-examination.

\section{Acknowledgment}
R.P. thanks Saurya Das and Tan Hai Siong and Ng Wei Khim for clarifying discussions.

\section*{Appendix A: The infinite well as the limit of a concave well}

\subsection{Power Law Potentials}
Consider the power-law potential 
\begin{eqnarray}
V(X)&:=& {\epsilon} \left({X \over L/2}\right)^{2\sigma} \label{power}
\end{eqnarray}
where $0<\sigma < \infty$, and $L,\epsilon$ are positive parameters.  As $\sigma \to \infty$ the potential describes an infinite well of width $L$.  
It is convenient in this case to study the problem in the momentum representation (\ref{momrep1},\ref{momrep}) which allows us to treat the infinite well as a limit of other smooth potentials.

Defining 
\begin{equation}
z^2 = E^{sc} = P^2 + \epsilon (2X/L) ^{2\sigma} \, , \nonumber 
\end{equation}
gives
\begin{equation}
X = (L/2) \epsilon^{-1/2\sigma} (z^2 - p^2)^{1/2\sigma} \label{power2}
\end{equation}
from which the canonical coordinate $x=X/f$ follows \cite{spectra}.

The phase-space area in the Sommerfeld-Wilson quantization rule is given by
\begin{eqnarray}
\oint x\ dp=  2\pi \bigl(n+ \gamma \bigr)  \label{BS}
\end{eqnarray} 
with $n$ integral; the constant $\gamma$ is  determined by the boundary conditions. Explicitly
\begin{eqnarray}
\oint x\ dp &=& 2 (L/2) \epsilon^{-1/2\sigma} \int_{-z}^{z}\frac{(z^2-p^2)^{1/2\sigma}}{\bigl[1- 2\alpha p + q \alpha^2 p^2 \bigr]} dp \label{semi}
\end{eqnarray}
where $\pm z$ are the classical turning points. In Ref.\cite{spectra} we discussed the integral (\ref{semi}) perturbatively in $\alpha$. Here we evaluate the integral exactly in the limit $\sigma \to \infty$. Since in that limit one approaches the hard boundary of the infinite well, we set $\gamma=0$ and $n=1,2,3...$. Thus one obtains

\begin{equation}
{2\pi n \over L} = \int_{-z}^{z}\frac{1}{\bigl[1- 2\alpha p + q \alpha^2 p^2 \bigr]} dp \, . \label{scexact} 
\end{equation}
The integral on the right is easily evaluated and the final results for the semi-classical energy agree with the exact results displayed in Sect(4). For the Snyder/Anti-Snyder cases one can simply set $\alpha \to 0$ and $q \alpha^2 \to \pm \beta$, obtaining again semi-classical expressions  that agree with the exact results of Sect.(4).  
  
One may also evaluate (\ref{scexact}) for small $\alpha p$ by expanding the integrand in (\ref{scexact}) for $\alpha z \ll 1$, to obtain \cite{spectra}
\begin{eqnarray}
\Delta E_{n}^{sc}&=& { 2 (q-4) \alpha^2 \pi^4 n^4  \over  3 L^4} + O(\alpha^4) \, . 
\end{eqnarray}

\section*{Appendix B: Hermiticity and Conservation of Probability Current}

In this section we discuss the relationship between the Hermiticity of the Hamiltonian and the deformed probability current for the third-order Schroedinger equation (\ref{6}).  Recall that for an operator $H$ to satisfy Hermiticity condition we must have
\begin{eqnarray}
\left[\int_{0}^{L} dx\ \psi_{j}^{*} (H\psi_{i})\right]^{*}&=& \int_{0}^{L}dx\ \psi_{i}^{*}H\psi_{j}\label{Herm1}
\end{eqnarray}
\noindent where $H:=H_{0}+H_{1}=p_{0}^2 +V(x)-2\alpha p_{0}^{3}$ and $(\psi_{i}, \psi_{j})$ are wave function for any states in the physical Hilbert space. 

Firstly, consider the undeformed part of the Hamiltonian (with $V(x)=0$ inside an infinite well). We obtain 
\begin{eqnarray}
\left[\int_{0}^{L} dx\ \psi_{j}^{*} (H_{0}\psi_{i})\right]^{*}&=&\left[-\int_{0}^{L} dx\ \psi_{j}^{*}\frac{\partial^2}{\partial x^2}\psi_{i}\right]^{*}\nonumber\\
&=&\int_{0}^{L}dx\ \psi_{i}^{*}\left(-\frac{\hbar^2}{2m}\frac{\partial^2}{\partial x^2}\right)\psi_{j}+ B_{0}\Bigr|_{0}^{L}\nonumber\\
&=&\int_{0}^{L}dx\ \psi_{i}^{*}H_{0}\psi_{j}+B_{0}\Bigr|_{0}^{L} \label{Herm2}
\end{eqnarray}
where we have performed and integration by parts twice and defined the boundary terms through the matrix
\begin{eqnarray}
B_{0}&:=&\Bigl[\psi_{i}^{*}\frac{\partial}{\partial x}\psi_{j}-\psi_{j}\frac{\partial}{\partial x}\psi_{i}^{*}\Bigr].
\end{eqnarray}

In order for the undeformed Hamiltonian to satisfy \eqref{Herm1}, we need the boundary terms to vanish and this 
can be satisfied by the usual Dirichlet conditions on the wavefunctions, see Sect.(4.2). 
Notice that the diagonal part of $B_{0}$ (i.e. $i=j$) is just the usual probability current density
\begin{eqnarray}
J_{0}(x)&:=&\frac{1}{i}\Bigl[\psi^{*}\frac{\partial}{\partial x}\psi-\psi\frac{\partial}{\partial x}\psi^{*}\Bigr]
\end{eqnarray}
which is required to vanish at each boundary if the particle is confined to the well. 

For $H_{1}$ Hermiticity requires
\begin{eqnarray}
\left[\int_{0}^{L} dx\ \psi_{j}^{*} (H_{1}\psi_{i})\right]^{*}&=&\left[-2\alpha\int_{0}^{L} dx\ \psi_{j}^{*}\frac{i\partial^3}{\partial x^3}\psi_{i}\right]^{*}\nonumber\\
&=&\int_{0}^{L}dx\ \psi_{i}^{*}\left(-2i\alpha\frac{\partial^3}{\partial x^3}\right)\psi_{j}+B_{1}\Bigr|_{0}^{L}\nonumber\\
&=&\int_{0}^{L}dx\ \psi_{i}^{*}H_{1}\psi_{j}+B_{1}\Bigr|_{0}^{L} \label{Herm3}
\end{eqnarray}
where we again performed integration by parts. The boundary matrix $B_{1}$ is
\begin{eqnarray}
B_{1}&=& 2i\alpha \left[\psi_{j}\frac{\partial^2 \psi_{i}^{*}}{\partial x^2}-\frac{\partial\psi_{j}}{\partial x}\frac{\partial\psi_{i}^{*}}{\partial x}+\left(\frac{\partial^2 \psi_{j}}{\partial x^2}\right)\psi_{i}^{*}\right]  \, .\nonumber\\
\end{eqnarray}  

Hence for the total Hamiltonian we need to impose the vanishing of the surface terms given by,
\begin{eqnarray}
B_{T}\Bigr|_{0}^{L} &=&  (B_0 + B_1)\Bigr|_{0}^{L}  \, .
\end{eqnarray}
Taking the diagonal term of $B_T$ we obtain the probability current density
\begin{eqnarray}
J(x)&=&J_{0}(x)+J_{1}(\alpha, x)\nonumber\\
&=&\left[\frac{1}{i}\Bigl(\psi^{*}\frac{\partial}{\partial x}\psi-\psi\frac{\partial}{\partial x}\psi^{*}\Bigr)+ 2\alpha\Bigl(\frac{\partial^2 |\psi|^2}{\partial x^2}-3\frac{\partial\psi}{\partial x}\frac{\psi^{*}}{\partial x}\Bigr)\right].\label{current1}
\end{eqnarray}
Notice that unlike the case of usual quantum mechanics, it is now not sufficient to just impose Dirichlet boundary conditions on the wavefunction to confine the particle in the well; one must either also impose vanishing of the slope of the wavefunction at each boundary or the vanishing of the total current at each boundary.

\end{document}